\documentclass[aps,prd,reprint,groupedaddress,showpacs,preprintnumbers]{revtex4-1}
\usepackage{graphicx}
\usepackage{amssymb,amssymb}
\usepackage{lineno}
\usepackage{subfigure}
\usepackage{hyperref} 

\usepackage[compact]{titlesec}
\titlespacing{\section}{0pt}{8pt}{4pt}
\titlespacing{\subsection}{0pt}{8pt}{4pt}
\titlespacing{\subsubsection}{0pt}{8pt}{4pt}

\newlength{\figwidth}
\begin{document}





\def\integLumiPrev{36~pb$^{-1}$}
\def\integLumi{1.0~fb$^{-1}$}
\def\integLumiMax{799~pb$^{-1}$}
\def\integLumiUncertainty{$\pm 11$\%}
\def\JERuncertainty{14} 
\def\MjjCut{200} %
\def\MjjLowerExclusion{400} %
\def\MjjUpperFitRange{1900} %
%
\def\MjjLowerLimitCTEQsixLone{2.06} 
\def\MjjLowerLimitCTEQsixLoneEXPECTED{2.01} 
\def\MjjLowerLimitStatOnlyCTEQsixLone{2.19} 
\def\MjjLowerLimitStatOnlyCTEQsixLoneEXPECTED{2.07} 
%
\def\MjjLowerLimitCTEQfiveL{2.14} 
\def\MjjLowerLimitCTEQfiveLEXPECTED{2.06} 
\def\MjjLowerLimitStatOnlyCTEQfiveL{2.26} 
\def\MjjLowerLimitStatOnlyCTEQfiveLEXPECTED{2.12}

\def\MjjLowerLimitMRST{2.99} 	
\def\MjjLowerLimitMRSTEXPECTED{2.81}          
\def\MjjLowerLimitStatOnlyMRST{3.07}           
\def\MjjLowerLimitStatOnlyMRSTEXPECTED{2.89}    

\def\MjjLowerLimitMRSTAxigluon{3.32}          
\def\MjjLowerLimitMRSTEXPECTEDAxigluon{3.07}   
\def\MjjLowerLimitStatOnlyMRSTAxigluon{3.33}    
\def\MjjLowerLimitStatOnlyMRSTEXPECTEDAxigluon{3.15}  

\def\MjjLowerLimitCTEQs8{1.92} 
\def\MjjLowerLimitCTEQEXPECTEDs8{1.77} 
\def\MjjLowerLimitStatOnlyCTEQs8{2.20} 
\def\MjjLowerLimitStatOnlyCTEQEXPECTEDs8{2.17} 

\def\MjjLowerLimitMRSTQBH{3.67} 
\def\MjjLowerLimitMRSTEXPECTEDQBH{3.64} 
\def\MjjLowerLimitStatOnlyMRSTQBH{3.68} 
\def\MjjLowerLimitStatOnlyMRSTEXPECTEDQBH{3.66}

\def\FchimjjLambda{6.12}  
\def\FchimjjEXPECTEDLambda{7.XX}  

\def\FchimjjQstar{3.00}
\def\FchimjjEXPECTEDQstar{3.XX}

\def\FchimjjQBH{3.62}
\def\FchimjjEXPECTEDQBH{3.XX}

\def\FchiKFacQCDCILambda{6.8}  
\def\FchiKFacQCDCIEXPECTEDLambda{5.2} 

\def\FchiQBH{3.69}
\def\FchiEXPECTEDQBH{3.37}
\def\FchiLowerLimitMRSTQBHnsix{3.69}
\def\FchiLowerLimitMRSTEXPECTEDQBHnsix{3.37}

\def\ElevenBinChiEXPECTEDLambda{7.31}
\def\ElevenBinChiEXPECTEDLambdaSigPlus{7.88}
\def\ElevenBinChiEXPECTEDLambdaSigMinus{6.93}
\def\ElevenBinChiLambda{6.65}

\def\ElevenBinChiEXPECTEDQBH{4.05}
\def\ElevenBinChiEXPECTEDQBHSigPlus{4.17}
\def\ElevenBinChiEXPECTEDQBHSigMinus{3.88}
\def\ElevenBinChiQBH{3.91}

\def\MjjLowerExclusion{0.60}
\def\HighestDijetMass{4}

\def\NpseudoExperiments{$10^3$}
\def\NpseudoExperimentsQCDfit{$10^4$}
\def\triggerThreshold{15} 
\def\pTcut{80} 
\def\topbottomLabel{~}
\def\leftrightLabel{~}
\def\Pythia{{\sc Pythia}}
\def\BlackMax{{\sc BlackMax}}
\def\Geant{{\sc Geant4}}
\def\CalcHEP{{\sc CalcHEP}}
\def\NLOJET{{\sc NLOJET++}}
\def\BumpHunter{{\sc BumpHunter}}
\def\TailHunter{{\sc TailHunter}}
\def\GeVcc{${\rm GeV}$}
\def\TeVcc{${\rm TeV}$}
\def\GeVc{${\rm GeV}$}
\def\invpb{${\rm pb}^{-1}$}
\def\mFchi{F_{\chi}}
\def\Fchi{$\mFchi$}
\def\mFchimjj{F_{\chi}(m_{jj})}
\def\Fchimjj{$\mFchimjj$}
\def\mjj{m_{jj}}
\def\qstar{{q^\ast}}
\def\pT{$p_{\rm T}$}
\def\mpT{p_{\rm T}}
\def\Acc{{\cal A}}   
\def\thetanp{\theta_{np}}


\preprint{CERN-PH-EP-2011-127 - Submitted to Physics Letters B}

\title{Search for New Physics in the Dijet Mass Distribution using 
1~fb$^{-1}$ of $pp$ Collision Data at $\sqrt{s}=7$~TeV collected by the ATLAS Detector 
}


\author{ATLAS Collaboration}
\affiliation{}





\begin{abstract}
Invariant mass distributions of jet pairs (dijets) produced in
LHC proton-proton collisions at a centre-of-mass energy $\sqrt{s}=7$~TeV
have been studied using a data set corresponding to an integrated luminosity
of \integLumi\ recorded in 2011 by ATLAS.
Dijet masses up to $\sim \HighestDijetMass$~TeV are observed in the data, and  
no evidence of resonance production over background is found.
Limits are set at 95\% CL for several new physics hypotheses:  
excited quarks are excluded for masses below \MjjLowerLimitMRST\ TeV,  
axigluons are excluded for masses below \MjjLowerLimitMRSTAxigluon\ TeV, 
and colour octet scalar resonances are excluded for masses below \MjjLowerLimitCTEQs8\ TeV.

\end{abstract}


\maketitle



\section{Introduction}

The Standard Model (SM) description of high energy proton-proton ($pp$) collisions
is based on the framework of quantum chromodynamics (QCD) in the perturbative regime,
where the most energetic collisions result from the 2 $\to$ 2 scattering of a pair
of partons (quarks or gluons).  Partons emerging from the collision shower and hadronise,
in the simplest case producing two jets of particles, a ``dijet'', that may
be reconstructed to determine the dijet invariant mass, $\mjj$, the mass of the
two-parton system. 

Previous studies of dijet mass distributions
\cite{Arnison1986244,Bagnaia1984283,CDF:2009DijetSearch,ATLAS:2010bc,
CMS:2010dijetmass,Aad:2011aj}\ have shown that these analyses are sensitive
to the highest mass scales accessible with hadronic final states.
In the present study, the dijet mass distribution is examined in a search for
resonances due to new phenomena localised near a given mass,
employing a data-driven background estimate that does not rely on detailed
QCD calculations.

In addition to new physics benchmarks used in previous ATLAS dijet analyses,
namely excited quarks ($\qstar$) \cite{Baur:1987ga,Baur:1989kv},
and axigluons ~\cite{Frampton1987a,Frampton1987b,Bagger1988},
the present study includes a third hypothetical object:
the colour octet scalar (s8), one of many
possible exotic colour resonances \cite{Han:2010rf}.
Any of these objects could produce a peak in the dijet spectrum in
the vicinity of their mass.

The present study is based on $pp$\ collisions at a centre-of-mass (CM)
energy of 7~TeV produced at the CERN Large Hadron Collider (LHC),
measured by the ATLAS detector. This data set corresponds to an integrated
luminosity of \integLumi\ recorded between March and June 2011.
The most stringent limits set previously by the ATLAS Collaboration
were based on the full 2010
data sample, corresponding an integrated luminosity of \integLumiPrev~\cite{Aad:2011aj}.
Excited quarks were excluded below 2.15 TeV, and axigluons below 2.10 TeV.
The CMS Collaboration has recently completed a dijet resonances analysis in 
\integLumi\ of 2011 data, excluding excited quarks below 2.49 TeV and axigluons below 2.47 TeV,
along with other limits \cite{CMS:Res2011}.

A detailed description of the ATLAS detector is available in~\cite{DetectorPaper}.  
The detector is instrumented over almost
the entire solid angle around the $pp$\ collision point with layers of
tracking detectors, calorimeters, and muon chambers.  
Jet measurements are made using a finely segmented calorimeter system designed
to detect the high energy jets that are the focus of this study with high
efficiency and excellent energy resolution.
ATLAS has a three-level trigger system, with the first level trigger (L1)
being based on custom-built hardware and the two higher level triggers (HLT)
being realised in software. 


ATLAS uses a right-handed coordinate system with the $z$-axis along the beam pipe.
The $x$-axis points to the centre of the LHC ring, and the $y$-axis points upward.
Cylindrical coordinates ($r$,$\phi$) are used in the transverse plane, $\phi$
being the azimuthal angle.  The pseudorapidity is defined in terms of the
polar angle $\theta$ as $\eta \equiv \mathrm{-ln\ tan}(\theta/2)$. 
Transverse momentum and energy are defined as $p_T = p$\,sin$\theta$
and $E_T = E$\,sin$\theta$, respectively.

The dijet mass, $\mjj$, is derived from the vectorial sum of the 
four-momenta of the two highest \pT\ jets in the event.
Kinematic criteria based on momentum and angular variables are applied
to increase the sensitivity to centrally produced high mass resonances.

The angular distribution for $2\to2$ parton scattering is predicted
by QCD in the CM frame of the colliding partons, which moves along the beamline
due to the differing momentum fractions (Bjorken $x$) of the colliding partons.
If $E$\ is the jet energy and $p_z$\ is the $z$-component of the
jet's momentum, the rapidity of the jet is given by
$y \equiv \frac{1}{2}\mathrm{ln}(\frac{E + p_z}{E - p_z})$.
The rapidities of the two highest \pT\ jets are denoted by $y_1$\ and $y_2$,
and the corresponding rapidity of these partons in their mutual
CM frame is $y^* = \frac{1}{2}(y_1 - y_2)$.

\section{Jet reconstruction and event selection}

Individual jets are reconstructed using the anti-$k_t$ jet clustering 
algorithm~\cite{antikT,Cacciari:2006} with the distance parameter $R = 0.6$.
The inputs to this algorithm are clusters~\cite{ATLAS-LARG-PUB-2008-002} of calorimeter cells with
energy depositions significantly above the measured noise.  
Jet four-momenta are constructed as the vectorial sum of
clusters of cells, treating each cluster as an ($E$, $\vec{p}$) four-vector with
zero mass, assuming that the corresponding particle stems from the primary vertex.  

The jet four-momenta are then corrected 
~\cite{ATLAS-CONF-2011-032}
as a function of $\eta$\ and \pT\ for various effects, the largest of which are
the hadronic shower response and detector material distribution.  
This is done using a calibration scheme based on Monte Carlo (MC) studies including full detector
simulation, and validated with extensive test-beam~\cite{Adragna:2010zz}
and collision data \cite{JES,JES2,ATLAS:JES3} studies.
Measured dijet mass distributions are not corrected for detector resolution,
which, in terms of mass smearing, is $\frac{\sigma_{m_{jj}} } {m_{jj}}
\simeq $5\% at $m_{jj}\simeq$1 TeV, drops to 4.5\% at 2 TeV, and
asymptotically approaches 4\% at $m_{jj}$ of 5 TeV and above.

The event selection starts with the first-level
   trigger, which selects events that have at least 
one large transverse energy deposition in the calorimeters, with the
transverse energy threshold increasing over the period of the data-taking
as the instantaneous luminosity of the LHC $pp$\ collisions increased.

To achieve the highest possible effective integrated luminosity, the current
data set has been recorded using a jet trigger that was usually
not prescaled. 
The chosen trigger has a nominal jet \pT\ threshold of 180 GeV.  
After applying all other analysis cuts, $\mjj$ is required
to be greater than 717 GeV in order to attain a trigger efficiency of at least 
99\% over the full range of the dijet mass distribution.

Events are required to have a primary collision vertex defined
by at least five charged-particle tracks. Events with a poorly measured
jet \cite{JetEtMiss_Cleaning_Note} with \pT\ greater than 30\% of the 
\pT\ of the next-to-leading jet are vetoed, to avoid cases where such a jet would
cause incorrect identification of the two leading jets.  
This rejects less than 0.002\%\ of the events.

Additional kinematic criteria are applied, 
requiring that the two leading jets each satisfy $|\eta_j| < 2.8$\ 
and that the rapidity in the parton CM frame satisfies $|y^*|< 0.6$.
These criteria favour central collisions and have been shown, based on studies
of expected signals and QCD background, to optimise the analysis sensitivity.


A final selection is made to avoid the calorimeter region from -0.1 to 1.5 in $\eta$ and from -0.9 to -0.5 in $\phi$,
which was in large part affected by readout problems for most of the data used 
in these studies. Events with jets in this region are discarded. This requirement
reduces the data set by 3.7\%.

\section{Comparing data to a smooth background}

\begin{figure}[!htb]
  \centering \includegraphics[width=0.45\textwidth]{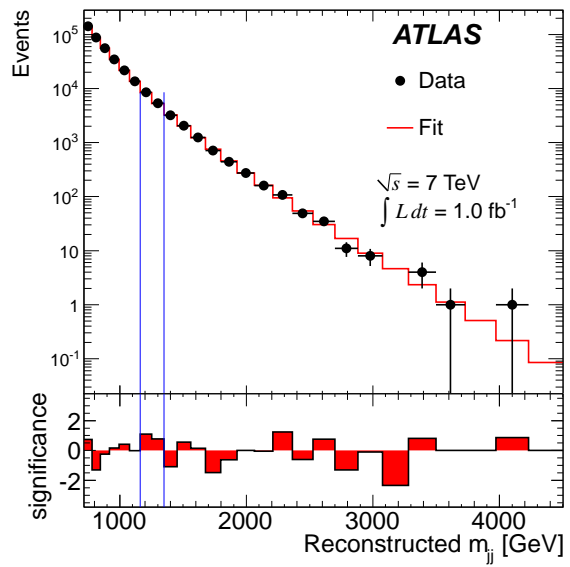}
  \caption{The reconstructed dijet mass distribution (filled points) fitted with
  a smooth functional form describing the QCD background.
  The bin-by-bin significance of the data-background difference is shown in the lower panel.
  Vertical lines show the most significant excess found by the \BumpHunter\ algorithm (see text).   \label{fig:1}}
\end{figure}

The observed dijet mass distribution after all selection cuts is shown in Fig.~\ref{fig:1}.
As in the previous ATLAS studies, the $\mjj$ spectrum is fit to the smooth functional
form 
\begin{equation}
  f(x) = p_1 (1 - x)^{p_2} x^{p_3 + p_4\ln x},
\end{equation}
where $x \equiv \mjj/\sqrt{s}$\ and the $p_i$\ are fit parameters.  
This ansatz has been shown empirically to accurately model the steeply falling QCD dijet 
mass spectrum~\cite{CDF:2009DijetSearch,ATLAS:2010bc,CMS:2010dijetmass,Aad:2011aj}. 
The $\mjj$ bins are of variable width,
increasing from $\sim50$\ to $\sim200$~GeV for dijet masses from 0.85 to 4.5~TeV, respectively,
to optimise the performance of the resonance search algorithm discussed in the next section.

The bottom plot of Fig.1 shows the significance, in standard deviations, 
of the difference between the data and the prediction in each bin.
These are purely statistical, and based on Poisson distributions.
The contents of a given bin are used to determine the $p$-value - the probability
of the background fluctuating higher than the observed excess, or lower than the observed
deficit.  The $p$-value is transformed to a significance, in terms of
an equivalent number of standard deviations (the $z$-value).
Where there is an excess (deficit) in data in a given bin, the significance is plotted as
positive (negative).
In mass bins with small expected number of events, where the observed number of events is similar
to the expectation, the Poisson probability of a fluctuation at least as high (low) as the observed
excess (deficit) can be greater than 50\%, as a result of the asymmetry of the Poisson distribution.
Such bins present no statistical interest and, for simplicity, bars are not drawn for them.

To determine the degree of consistency between data and the fitted background, the $p$-value
of the fit is obtained by calculating the $\chi^2$ from the data, and comparing this result to the
$\chi^2$ distribution obtained from pseudoexperiments.  The resulting $p$-value is 0.96, 
showing that there is good agreement between the data and the functional form.

\section{Search for resonances}

As a more sensitive test, the 
\BumpHunter\ algorithm~\cite{Aaltonen:2008vt,Choudalakis:2011bh}
is used to establish the presence or absence of a resonance in the dijet mass spectrum.
To optimise the sensitivity of this algorithm, the $\mjj$ binning strategy
is to establish a minimum width for resonances to be considered physical.
To this end, the relatively narrow $\qstar$ $\mjj$ template from full
MC simulation~\cite{ATLSIM}, described below for subsequent studies,
has been used to establish the binning.  If the width of the resonance is defined
as $\pm 1 \sigma$, the greatest sensitivity at the minimum width is achieved
by setting the bin width to $1 \sigma$, half the resonance width.
The final result of this procedure is that the variable bin sizes
are typically 6.5\% to 7.0\% of $\mjj$ in width, somewhat wider than detector resolution due to the finite natural width of $\qstar$, which varies between about 3\% and 3.5\% of the $\qstar$ mass.

In the current implementation, the \BumpHunter\ algorithm searches for the signal
window with the most significant excess of events above background.  
Starting with a two-bin window, the algorithm
increases the signal window and shifts its location until all possible bin ranges,
up to half the mass range spanned by the data, have been tested.  
The most significant departure from the smooth spectrum (``bump'') is defined by the
set of bins that have the smallest probability of arising from a background
fluctuation assuming Poisson statistics.

The \BumpHunter\ algorithm accounts for the so-called ``look elsewhere effect'' 
(or ``trials factor effect'')~\cite{lookelsewhere}
by performing a series of pseudoexperiments to determine
the probability that random fluctuations in the background-only hypothesis
would create an excess as significant as the one observed anywhere in the spectrum.
Variable width binning reduces the penalty due to this effect,
while retaining sensitivity.

To prevent any new physics signal from biasing the background estimate,
if the biggest local excess from the background fit has a p-value smaller than 0.01,
this region is excluded and a new background fit is performed.  No such exclusion
is needed for this data set.

The most significant discrepancy identified by the \BumpHunter\ algorithm in the observed dijet mass distribution reported in Fig.~\ref{fig:1} is a 2-bin excess in the interval 1.16 to 1.35 TeV.  
The probability of observing such an excess or larger somewhere in the mass spectrum
for a background only hypothesis is 0.82.
This test shows that there is no evidence for a resonance signal in
the $\mjj$\ spectrum.

\section{New physics models}

Exclusion limits are set on three new physics scenarios
expected to give rise to resonant dijet production.

For the first of these, excited quarks, $\qstar$,
a $qg\to \qstar$ production model~\cite{Baur:1987ga,Baur:1989kv} is used,
with the assumption of spin 1/2 and quark-like SM coupling constants.
The compositeness scale ($\Lambda$) is set to the $\qstar$\ mass.
Signal events are produced using the \Pythia\ event generator~\cite{pythia},
a leading-order parton-shower MC generator, with the 
MRST2007LO*~\cite{Sherstnev:2007nd} parton distribution functions (PDF's),
with settings established by the 
ATLAS default MC10~\cite{ATLAS_MC10_PYTHIA_AMBT1} Monte Carlo tune.
The renormalization and factorization scales are set to the mean
\pT\ of the two leading partons for each event. \Pythia\ is also used to decay the excited quarks
to all possible SM final states, which are predominantly $qg$, but also $qW$,
$qZ$, and $q\gamma$. The generated events are passed through the detailed
simulation of the ATLAS detector~\cite{ATLSIM}, which uses the 
\Geant\ package~\cite{Agostinelli:2002hh} for simulation of particle 
transport, interactions, and decays.
The simulated events are then reconstructed in the same way 
as the data to produce predicted dijet mass distributions
that can be compared with the observed distributions. 

The second model is axigluon production~\cite{Frampton1987a,Frampton1987b,Bagger1988}\ 
via an interaction given by the Lagrangian
\begin{eqnarray}
\mathcal{L}_{Aq\bar{q}} = g_{QCD}\bar{q}A^{a}_{\mu}\frac{\lambda^a}{2} \gamma^{\mu}\gamma_{5}\ q,
\end{eqnarray}
where $g_{QCD}^2 = 4\pi\alpha_s$ is the QCD coupling constant and $A^{a}_{\mu}$ is the axigluon field representing
a massive state with axial coupling to quarks.  Parity conservation prevents the axigluon from
coupling to two gluons.
Parton-level events are generated, at leading-order approximation, using the \CalcHEP\ Monte Carlo package~\cite{Pukhov:2004ca},
for chosen masses, $m$, of the axigluon.  The MRST2007LO* PDF set was used.
The axigluon dijet mass has longer tails at high and low masses than the $\qstar$ distribution,
but these two shapes are interchangeable within the range 0.7$m$ to 1.3$m$ 
for all masses of interest.  Since the axigluon tails outside this range are well below the
SM background, the predicted signal may be analyzed by cutting events beyond this range and
accounting for the reduced acceptance.  The axigluon MC prediction for $\sigma\times {\cal A}$,
the production cross section within the acceptance,
is defined to include these cuts by applying them at the level of \CalcHEP\ generation, 
along with the kinematic cuts in \pT\ and rapidity.
In the limit setting analysis, these axigluon results are compared to the observed $\sigma\times {\cal A}$ 
limits from the $\qstar$ analysis.  
This method is discussed in more detail in Section~\ref{sec:limit}.

The third resonant hypothesis, the colour octet scalar (s8) model,
is a prototype for many possible exotic coloured resonances~\cite{Han:2010rf}.
Colour octet resonances can couple to gluons, which have large parton luminosity at the LHC. 
One possible interaction is 
\begin{eqnarray}
\mathcal{L}_{gg8}=g_{QCD}d^{ABC}\frac{\kappa_s}{\Lambda_s} S_{8}^{A} F^B_{\mu\nu}F^{C,\mu\nu},
\label{tensscal.EQ}
\end{eqnarray}
where $S_8^{A}$ is the colour octet scalar field, $\kappa_s$ is the scalar coupling (assumed to be unity),
and $d^{ABC}$ is the SU(3) isoscalar factor;
$\Lambda_s$ is the new physics scale which is set to the resonance mass, $M_{s_{8}}$. 
This model leads to a very simple event topology, with two gluons in the initial and final states,
 yielding high $p_T$ dijets. 
{\sc MadGraph}~5~\cite{Alwall:2011uj} is used to generate parton level events at leading-order approximation. 
\Pythia\ with {\sc CTEQ6L1} PDF's is used in this generation, with the ATLAS MC09'
tune~\cite{ATLAS_MC09P}.
These samples are processed through the full ATLAS detector simulation.

The observed limits on s8 are less strict than the corresponding $q^\ast$ limits,
in part because the s8 signal is much wider than $q^\ast$.  Much of this width increase
is due to final state radiation, which is larger for gluon-jets than for
quark-jets. In addition, the initial state for s8 production contains gluons,
which have small parton density at high mass.  Thus, s8 are much more likely
to be off-mass-shell than $q^\ast$.

\section{Model dependent limit setting} 
\label{sec:limit}

In the absence of any observed significant discrepancy from the zero-signal hypothesis,
the Bayesian method documented in \cite{Aad:2011aj} is used to set 95\% credibility-level
(CL) upper limits.

\begin{figure*}[!htb]
  \centering
    \subfigure[Excited-quark and axigluon models.]{
      \label{fig:reslimsiga}
      \includegraphics[width=0.45\textwidth]{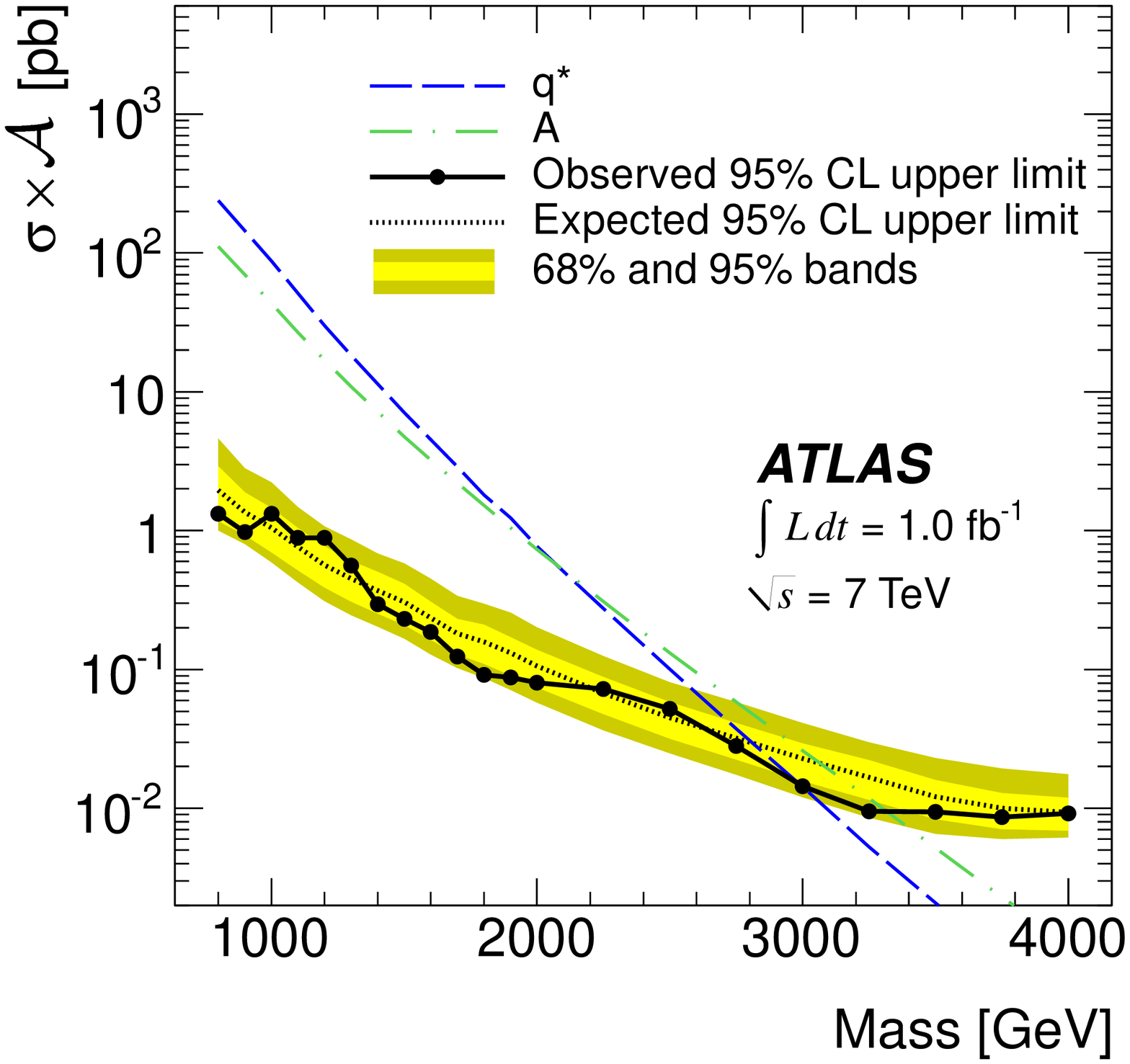}}
    \subfigure[Colour octet scalar model.]{
      \label{fig:mjjcolouroctet}
      \includegraphics[width=0.45\textwidth]{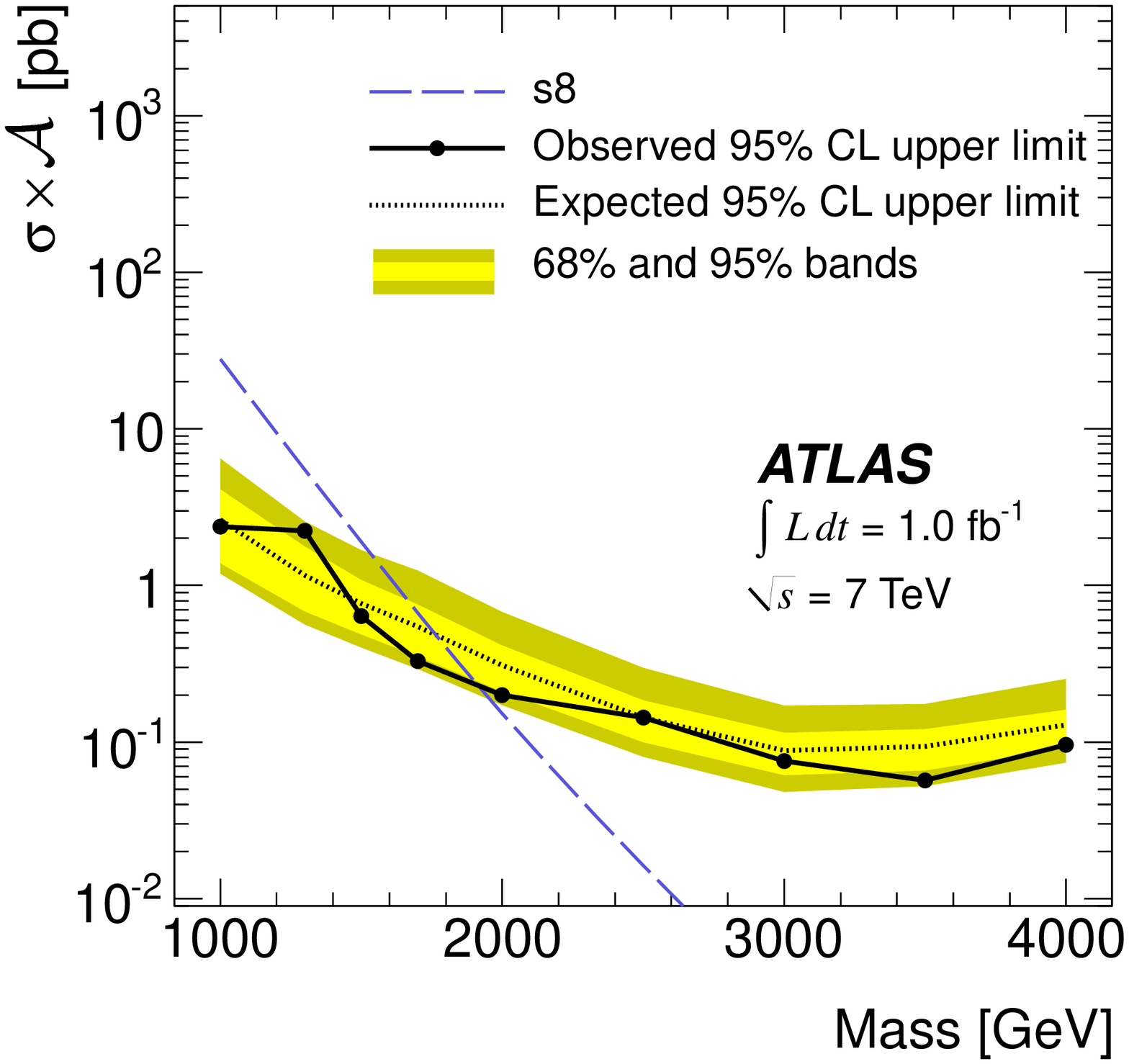}}
    \caption{The 95\% CL upper limits on $\sigma\times {\cal A}$ as a function of particle mass (black filled
		circles). The black dotted curve shows the 95\% CL upper limit expected from Monte Carlo and the 
		light and dark yellow shaded bands represent the 68\% and 95\% contours of the expected limit, respectively. 
		Theoretical predictions for $\sigma\times {\cal A}$ are shown in (a) for excited quarks
        (blue dashed) and axigluons (green dot-dashed), and in (b) for colour octet scalar resonances (blue dashed).
		For a given new physics model, the observed (expected) limit occurs at the crossing of its
        $\sigma\times {\cal A}$ curve with the observed (expected) 95\% CL upper limit curve.}
  \label{fig:combinedfig}
\end{figure*}

Bayesian credibility intervals are set by defining a posterior probability density from the
Poisson likelihood function for the observed mass spectrum, obtained by a fit to the background
functional form and a signal shape derived from MC simulations.
A prior probability density constant in all positive values of signal cross section,
and zero at negative values, is used.
The posterior probability is then integrated to determine the
95\% CL for a given range of models, usually parameterised by the mass of the
resonance.

Limits are determined on $\sigma\times {\cal A}$ for a hypothetical new particle 
decaying into dijets. The acceptance includes all reconstruction steps and analysis cuts described 
above, and assumes that the trigger is fully efficient. (The efficiency is greater than 99\% 
for all analyses.)

The effects of systematic uncertainties due to the knowledge of the luminosity and of the jet energy scale (JES) 
are included. The luminosity
uncertainty for the 2011 data is 3.7\% ~\cite{ATLAS-CONF-2011-116}.
The systematic uncertainty on the JES is taken from the 2010 data~\cite{ATLAS-CONF-2011-032} analysis,
and is adapted to the 2011 analysis taking into account in particular the new event pileup conditions
(described below).  The JES uncertainty shifts resonance peaks by less than 4\%.
The background parameterization uncertainty is taken from the fit results, as described in~\cite{Aad:2011aj}.
The effect of the jet energy resolution (JER) uncertainty is found to be negligible.  
All of these uncertainties are incorporated into the fit by varying all
sources according to Gaussian probability distributions and convolving them with the Bayesian
posterior probability distribution. Credibility intervals are then calculated numerically from the
resulting convolutions.  No uncertainties are associated with the theoretical model
of new physics, as in each case the model is a benchmark that incorporates a specific choice of model parameters, of PDF set, and of MC tune.
Previous ATLAS studies have already explored the impact of different MC tunes and PDF sets on the $q^\ast$ theoretical prediction \cite{ATLAS:2010bc}.

In 2011, the instantaneous luminosity has risen to a level where corrections
must be made for multiple $pp$ collisions occurring in the same
bunch crossing (``pileup''), whose presence affects the measurement of calorimeter energy
depositions associated with the hard-scattering event under study.
All simulated samples used in this analysis include a Poisson distributed number of MC minimum bias events added
to the hard interaction to account for ``in-time'' pileup caused by additional collisions in
the same bunch crossing.  Further account must be taken of ``out-of-time'' pileup
originating from collisions in bunches preceding or following the one of interest, due to the long response time of the liquid argon calorimeters.
With the 50 ns bunch spacing in the LHC for these data, up to 12 preceding bunches
and 1-2 following bunches contribute to out-of-time pileup.
Although the conditions modelled in MC are realistic, they may not perfectly match the data
due to bunch train structure and instantaneous luminosity variations
in the LHC. The MC events are therefore reweighted to remove these 
residual differences.
Following this procedure the pileup description in MC is sufficiently good
that no additional uncertainty on the JES is required for jets with \pT\ $> 100$ GeV.

The resulting limits are shown in Fig.~\ref{fig:combinedfig}. For excited quarks,
the acceptance
$\cal A$ ranges from 37 to 51\% for $m_\qstar$ 
varying from 0.8 to 5.0 TeV, and is never lower than 47\% above masses of 1.1~\rm TeV. 
The main impact on the acceptance comes from the rapidity requirements. 
Using the theoretical prediction for $\qstar$ production described above, the expected mass limit at 95\% CL
is \MjjLowerLimitMRSTEXPECTED\ TeV, and the observed limit is \MjjLowerLimitMRST\ TeV.

The axigluon results are obtained from the $\sigma\times {\cal A}$ limits determined 
from the $\qstar$ analysis.  
The axigluon theoretical prediction is derived from the cross section provided by \CalcHEP\,
at each simulated mass, $m$, within the restricted mass range 0.7$m$ to 1.3$m$, 
after applying the kinematic selections. 
Using the axigluon theoretical $\sigma\times {\cal A}$ thus defined, the expected axigluon mass
limit at 95\% CL is \MjjLowerLimitMRSTEXPECTEDAxigluon\ TeV, 
and the observed limit is \MjjLowerLimitMRSTAxigluon\ TeV. 
This method has been confirmed by full simulation of axigluon samples at three
mass points, showing that the differences between parton level and full simulation 
are negligible compared to the effects of other uncertainties.

Figure~\ref{fig:mjjcolouroctet} shows the limits on the accepted cross section $\sigma\times {\cal A}$ for
colour octet resonances.
The expected mass limit at 95\% CL is 
\MjjLowerLimitCTEQEXPECTEDs8\ TeV, and the observed limit is \MjjLowerLimitCTEQs8\ TeV.
Since the colour octet scalar cross section decreases much more rapidly with $m$
than those for excited quark and axigluon production,
the resulting limits are considerably lower.

For all three models used in these studies, if systematic uncertainties had not been
included the exclusion limits would be approximately 60 GeV higher.

\section {Model independent limit setting}


In addition to specific theoretical models, limits are set to a collection of hypothetical signals that are assumed to be Gaussian-distributed in $\mjj$ with mean ($m_{\text{G}}$) ranging from 0.9 to 4.0 TeV and standard deviation ($\sigma_{\text{G}}$) from 5\% to 15\% of the mean.

Systematic uncertainties are treated using the same methods as applied in
model dependent limit setting.  The only difference for the Gaussian
analysis arises from the decay of the dijet final state not being simulated.
In place of this, it is assumed that the dijet signal distribution
is Gaussian in shape, and the JES is adjusted by modelling it as an uncertainty
of 4\% in the central value of the Gaussian signal.

\begin{figure}[!htb]
  \centering \includegraphics[width=0.45\textwidth]{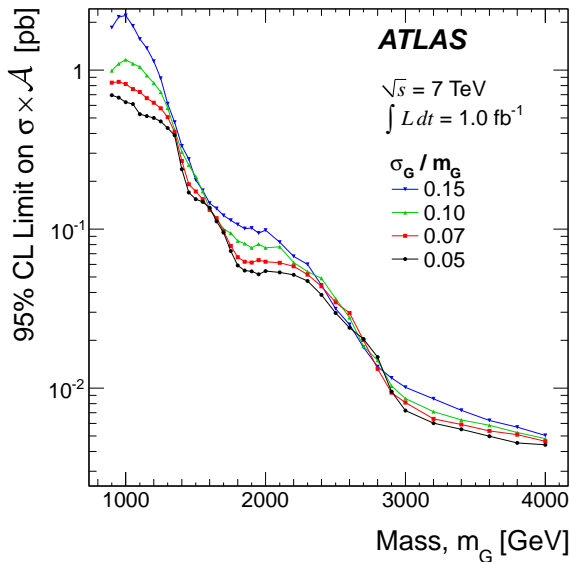}
  \caption{The 95\%\ CL upper limits on $\sigma\times {\cal A}$ for a simple Gaussian
           resonance decaying to dijets as a function of the mean mass, $m_{\text{G}}$, for
           four values of $\sigma_{\text{G}}/m_{\text{G}}$, taking into account
           both statistical and systematic uncertainties.}
  \label{fig:mjjGaussian}
\end{figure}

The resulting limits on $\sigma\times {\cal A}$ for the Gaussian
template
model are shown in Fig.~\ref{fig:mjjGaussian}.  Relative to previous 
studies~\cite{Aad:2011aj}\ they are substantially improved in the region above 900 GeV.
These results may be utilised to set limits on new physics models beyond
those considered in these studies, using the procedure described in the Appendix.

\section{Conclusion}


The dijet mass spectrum measured by the ATLAS experiment has been examined in a search for resonances from new phenomena,
using \integLumi\ of 7 TeV $pp$ collision data taken in 2011.
The observed distribution, which extends up to masses of
$\approx \HighestDijetMass$~TeV, is in good agreement with a smooth function representing the SM expectation. No evidence
for the production of new resonances is found.
95\% CL mass limits using Bayesian methodology have been set in the context of several models of new physics, as summarized in Table~\ref{tab:Summary}.
For excited quarks and axigluons, the current results exceed the limits obtained by ATLAS with
the 2010 data by approximately one TeV. Exclusion limits on colour octet scalar resonances
have been established for the first time in ATLAS.
The limits reported in this paper are the most stringent to date.

\begin{table}[!hbt]
\centering {
\caption {The 95\% CL mass lower limits for the models of new physics examined in this study.
They have been obtained with Bayesian analyses and include systematic uncertainties.}
\label{tab:Summary}
\begin{tabular}{lcc}
\hline
Model \quad \quad &  \multicolumn{2}{c}{95\%\ CL Limits (TeV)} \\
            &  Expected  &  Observed\\
\hline 
   Excited Quark $\qstar$  &   \MjjLowerLimitMRSTEXPECTED         &  \MjjLowerLimitMRST     \\ 
   Axigluon                &   \MjjLowerLimitMRSTEXPECTEDAxigluon &  \MjjLowerLimitMRSTAxigluon \\  
   Colour Octet Scalar     &   \MjjLowerLimitCTEQEXPECTEDs8       &  \MjjLowerLimitCTEQs8 \\  
\hline
\end{tabular}
     } 
\end{table}

\newpage
{\bf Appendix: Setting limits on new models}
\medskip

The following procedure is appropriate for resonances that are approximately
Gaussian near the core, and with tails that are well below the background.
For convenience, the results of Fig.~\ref{fig:mjjGaussian} are provided in
Table ~\ref{qnum.gausslims}.

(1) For a MC sample generated with the mass of the hypothetical new particle
set to $M$, compute an initial acceptance including the branching ratio into dijets.
Then apply the kinematic cuts on the parton $\eta$, \pT, and $|y^*|$ used in this analysis.
(2) Approximate the reduction of acceptance due to the calorimeter (temporary) readout problem
by eliminating events where a parton enters the region -0.1 to 1.5 in $\eta$,
and -0.9 to -0.5 in $\phi$.  
(Indicatively, the acceptance of $q^\ast$ is reduced by a factor 0.92.)
(3) Smear the signal mass distribution to reflect the detector resolution.
In the absence of a better detector simulation tool, use the mass resolution
given in Section II, which is derived from full ATLAS simulation.
(4) Since a Gaussian signal shape has been assumed in determining the limits, 
any long tails in the reconstructed $\mjj$\ should be removed in the sample under study. 
The recommendation (based on optimization using $q^\ast$ templates) is to retain
events with $\mjj$\ between 0.8$M$ and 1.2$M$.
The mean mass, $m$, of this truncated signal should be calculated.  
(5) The fraction of MC events surviving the first four steps determines 
 the modified acceptance, ${\cal A}$.
(6) From Table ~\ref{qnum.gausslims} select $m_{\text{G}}$ so that $m_{\text{G}}=m$.
If the exact value of 
$m$ is not among the listed values of $m_{\text{G}}$, check the limit for the two values
of $m_{\text{G}}$ that are directly above and below
$m$, and use the larger of the two limits to be conservative.
(7) To retain enough of the information in the full signal template, and at the
same time reject tails that would invalidate the Gaussian approximation, the
following truncation procedure is recommended.
For this mass point, choose a value of $\sigma_{\text{G}}/m_{\text{G}}$
such that the width 2$\sigma_{\text{G}}$ is well contained in the (truncated) mass range.
For the $q^\ast$ a good choice is empirically found to be $\sigma_G = (1.2M-0.8M)/5$.
This $\sigma_G$ corresponds to a Gaussian distribution contained within the
truncation interval of $[0.8M,1.2M]$, since the interval $[0.8M,1.2M]$
corresponds to $[m_G -2.5\sigma_G , m_G +2.5\sigma_G]$.

For the $\qstar$ case a good choice is $\sigma_{\text{G}}$ =
(1.2$M$-0.8$M$)/5 so that 95\% of the Gaussian spans $4 \times(0.4/5)M$.
Use this value to pick the closest $\sigma_{\text{G}}$/$m_{\text{G}}$ value,
rounded up to be conservative.
(8) Compare the tabulated 95\% CL upper limit corresponding
   to the chosen $m_{\text{G}}$ and $\sigma_{\text{G}}/m_{\text{G}}$ values to the $\sigma\times {\cal A}$ obtained from the theoretical
   cross section of the model multiplied by the acceptance defined in step (5) above.

\begin{table}[!htb]
\caption{The 95\% CL upper limit on $\sigma\times {\cal A}$ [pb] for the Gaussian ``model-independent'' scenario.
     The symbols $m_{\text{G}}$ and $\sigma_{\text{G}}$ are, respectively, the mean mass and standard deviation of the Gaussian.}
\begin{center}
\begin{tabular}{|l|llll|}
\hline
 $m_{\text{G}}$ &          & $\sigma_{\text{G}}/m_{\text{G}}$ & & \\
 (GeV)  &  5\%   &  7\%   &  10\%  &  15\%  \\
\hline
 900   &  0.69  &  0.83  &  0.99  &  1.9  \\ 
 950   &  0.67  &  0.84  &  1.1  &  2.2  \\ 
 1000   &  0.63  &  0.82  &  1.2  &  2.2  \\ 
 1050   &  0.61  &  0.76  &  1.1  &  1.9  \\ 
 1100   &  0.53  &  0.73  &  1.0  &  1.6  \\ 
 1150   &  0.51  &  0.67  &  0.93  &  1.4  \\ 
 1200   &  0.50  &  0.62  &  0.83  &  1.1  \\ 
 1250   &  0.48  &  0.58  &  0.73  &  0.89  \\ 
 1300   &  0.43  &  0.51  &  0.58  &  0.61  \\ 
 1350   &  0.39  &  0.41  &  0.42  &  0.47  \\ 
 1400   &  0.24  &  0.27  &  0.31  &  0.33  \\ 
 1450   &  0.17  &  0.19  &  0.25  &  0.28  \\ 
 1500   &  0.15  &  0.17  &  0.21  &  0.20  \\ 
 1550   &  0.15  &  0.15  &  0.17  &  0.18  \\ 
 1600   &  0.14  &  0.13  &  0.13  &  0.15  \\ 
 1650   &  0.11  &  0.12  &  0.12  &  0.13  \\ 
 1700   &  0.095  &  0.097  &  0.100  &  0.12  \\ 
 1750   &  0.073  &  0.078  &  0.094  &  0.11  \\ 
 1800   &  0.059  &  0.067  &  0.084  &  0.11  \\ 
 1850   &  0.055  &  0.062  &  0.081  &  0.10  \\ 
 1900   &  0.054  &  0.062  &  0.076  &  0.10  \\ 
 1950   &  0.052  &  0.064  &  0.081  &  0.094  \\ 
 2000   &  0.054  &  0.062  &  0.076  &  0.098  \\ 
 2100   &  0.053  &  0.061  &  0.078  &  0.083  \\ 
 2200   &  0.052  &  0.058  &  0.062  &  0.067  \\ 
 2300   &  0.047  &  0.052  &  0.054  &  0.060  \\ 
 2400   &  0.039  &  0.044  &  0.049  &  0.044  \\ 
 2500   &  0.030  &  0.035  &  0.037  &  0.032  \\ 
 2600   &  0.024  &  0.030  &  0.028  &  0.025  \\ 
 2700   &  0.020  &  0.020  &  0.018  &  0.018  \\ 
 2800   &  0.016  &  0.013  &  0.015  &  0.014  \\ 
 2900   &  0.009  &  0.009  &  0.010  &  0.012  \\ 
 3000   &  0.007  &  0.008  &  0.009  &  0.010  \\ 
 3200   &  0.006  &  0.006  &  0.007  &  0.009  \\ 
 3400   &  0.005  &  0.006  &  0.006  &  0.007  \\ 
 3600   &  0.005  &  0.005  &  0.006  &  0.006  \\ 
 3800   &  0.005  &  0.005  &  0.005  &  0.006  \\ 
 4000   &  0.004  &  0.005  &  0.005  &  0.005  \\ 
\hline
\end{tabular}
\end{center}
\label{qnum.gausslims}
\end{table}

\bigskip
{\bf Acknowledgements}
\medskip

We thank CERN for the very successful operation of the LHC, as well as the
support staff from our institutions without whom ATLAS could not be
operated efficiently.

We acknowledge the support of ANPCyT, Argentina; YerPhI, Armenia; ARC,
Australia; BMWF, Austria; ANAS, Azerbaijan; SSTC, Belarus; CNPq and FAPESP,
Brazil; NSERC, NRC and CFI, Canada; CERN; CONICYT, Chile; CAS, MOST and
NSFC, China; COLCIENCIAS, Colombia; MSMT CR, MPO CR and VSC CR, Czech
Republic; DNRF, DNSRC and Lundbeck Foundation, Denmark; ARTEMIS, European
Union; IN2P3-CNRS, CEA-DSM/IRFU, France; GNAS, Georgia; BMBF, DFG, HGF, MPG
and AvH Foundation, Germany; GSRT, Greece; ISF, MINERVA, GIF, DIP and
Benoziyo Center, Israel; INFN, Italy; MEXT and JSPS, Japan; CNRST, Morocco;
FOM and NWO, Netherlands; RCN, Norway; MNiSW, Poland; GRICES and FCT,
Portugal; MERYS (MECTS), Romania; MES of Russia and ROSATOM, Russian
Federation; JINR; MSTD, Serbia; MSSR, Slovakia; ARRS and MVZT, Slovenia;
DST/NRF, South Africa; MICINN, Spain; SRC and Wallenberg Foundation,
Sweden; SER, SNSF and Cantons of Bern and Geneva, Switzerland; NSC, Taiwan;
TAEK, Turkey; STFC, the Royal Society and Leverhulme Trust, United Kingdom;
DOE and NSF, United States of America.

The crucial computing support from all WLCG partners is acknowledged
gratefully, in particular from CERN and the ATLAS Tier-1 facilities at
TRIUMF (Canada), NDGF (Denmark, Norway, Sweden), CC-IN2P3 (France),
KIT/GridKA (Germany), INFN-CNAF (Italy), NL-T1 (Netherlands), PIC (Spain),
ASGC (Taiwan), RAL (UK) and BNL (USA) and in the Tier-2 facilities
worldwide.

\newpage


\bibliography{DijetResonancePLB2011}

\clearpage
\begin{flushleft}
{\Large The ATLAS Collaboration}

\bigskip

G.~Aad$^{\rm 48}$,
B.~Abbott$^{\rm 111}$,
J.~Abdallah$^{\rm 11}$,
A.A.~Abdelalim$^{\rm 49}$,
A.~Abdesselam$^{\rm 118}$,
O.~Abdinov$^{\rm 10}$,
B.~Abi$^{\rm 112}$,
M.~Abolins$^{\rm 88}$,
H.~Abramowicz$^{\rm 153}$,
H.~Abreu$^{\rm 115}$
% ...
\bigskip

% ...
$^{10}$ Institute of Physics, Azerbaijan Academy of Sciences, Baku, Azerbaijan\\
$^{11}$ Institut de F\'isica d'Altes Energies and Departament de F\'isica de la Universitat Aut\`onoma  de Barcelona and ICREA, Barcelona, Spain\\
% ...
$^{48}$ Fakult\"{a}t f\"{u}r Mathematik und Physik, Albert-Ludwigs-Universit\"{a}t, Freiburg i.Br., Germany\\
$^{49}$ Section de Physique, Universit\'e de Gen\`eve, Geneva, Switzerland\\
% ...
$^{88}$ Department of Physics and Astronomy, Michigan State University, East Lansing MI, United States of America\\
% ...
$^{111}$ Homer L. Dodge Department of Physics and Astronomy, University of Oklahoma, Norman OK, United States of America\\
$^{112}$ Department of Physics, Oklahoma State University, Stillwater OK, United States of America\\
% ...
$^{115}$ LAL, Univ. Paris-Sud and CNRS/IN2P3, Orsay, France\\
% ...
$^{118}$ Department of Physics, Oxford University, Oxford, United Kingdom\\
% ...
$^{153}$ Raymond and Beverly Sackler School of Physics and Astronomy, Tel Aviv University, Tel Aviv, Israel\\
% ...
\end{flushleft}


\end{document}